\documentclass[a4paper,preprint]{aastex}

\usepackage{graphicx}

\def\be{\begin{equation}}

\def\ee{\end{equation}}

\begin{document}

\title{A Model of Void Formation}

\author{Yasmin Friedmann and Tsvi Piran\thanks{Racah Institute, The
    Hebrew University, Jerusalem, 91904, Israel;
yasmin@merger.fiz.huji.ac.il; tsvi@nikki.fiz.huji.ac.il}}

\begin{abstract}

We introduce a simple model for the formation of voids. In this
model the underdensity of galaxies in voids is the product of two factors.
The first arises from a gravitational expansion of
the negative density perturbation.  The second is due to
biasing: galaxies are less likely to form in an underdense region.
One feature of the model is an upper cutoff in void sizes. We
calculate the volume filling factor of characteristic voids for
different CDM models and find that our formation model points to
$\Lambda$CDM models as preferred models of the power spectrum.  A
natural consequence of our model is that the underdensity of the dark
matter inside voids is smaller than the galaxy underdensity.

\end{abstract}




\section{Introduction}

Visual inspection of redshift surveys has revealed
\citep{lapp,bootes87,gellhuch,costa94,CS,LCRS96} that a large fraction
of the universe is made of ''voids'': regions in which the typical
galaxy density is significantly lower than the mean galaxy density.
Most galaxies tend to be found in two dimensional sheets that
encompass these voids.  Using the ``VOID FINDER'' algorithm, an
automated algorithm that detects voids in three dimensional surveys
and measures their volume, El-Ad and Piran determined the sizes and
depths of voids in several surveys: SSRS2, IRAS and ORS \citep{hagai1,
hagaiiras, hagaiors}. They found that the void distribution is
remarkably stable and that different surveys that encompass the same
regions in the sky see the same voids. They also found that: (i) The
voids occupy $\approx 50\%$ of the volume.  (ii) Void radii are in the
range 13-30$h^{-1}$ Mpc.  There appears to be an upper cutoff to the
sizes of the voids.  This upper limit does not depend on the
properties of a particular survey or on the effective depth.  This
upper cutoff is also seen in visual inspection of the deepest survey
existing today, the LCRS \citep{LCRS96}, whose effective depth is
100$h^{-1}$ Mpc.  (iii) The density contrast of galaxies in voids is
in the range $[-0.70,-0.95]$.

In current surveys the sizes of observed structures are not much
smaller than the effective depth of the survey.  Therefore only about
a dozen voids have been identified so far.  Hence, there is not enough
data to produce a good statistics on the distribution of the voids'
sizes and depths.  This situation will change with the new generation
of automated redshift surveys - the 2dF \citep{lahav96} and the SDSS
\citep{loveday}. With these surveys we will be able to identify dozens
of voids and to quantify their features.  With this situation in mind
we are beginning to set here the ground for the analysis of these
properties and for a comparison of the observations with a simple
model for void formation.

The existence of significant inhomogeneities on the scale of tens of
Mpc should be an important clue as to the formation of large scale
structure, and can be useful in exploring the power spectrum on this
scale range. Indeed positive fluctuations on scales of 10 Mpc
(clusters and super clusters) provide a powerful tool to explore the
power spectrum (see \citet{bahcall}). However voids have not been used
as yet. This is due to several reasons. The present day sky surveys
are not comprehensive enough to allow a full quantitative assessment
of void sizes and of the distribution function of void sizes.
Secondly, there is a lack of a simple theory for the formation of
voids.

A theory for the formation of voids should explain the physical
mechanisms which operate in the formation of the voids. It should be
able to explain qualitatively the appearance of the apparent upper
cutoff on their sizes.  Using this theory one could compare the
properties of voids (more specifically, voids sizes and filling
factor) that arise from different primordial perturbation spectra with
the observations.  Another goal of such a model is to predict the
underdensity of the dark matter within the voids and to provide us
with a prediction of the effective biasing factor within the voids.

\citet{bdglp}, \citet{dub93} and \citet{pi97} considered a purely
gravitational scenario for the formation of voids.  Their model is
based on the assumption that light traces matter on the scale of
voids. In this case the observed underdensity in the galaxy
distribution corresponds to a comparable underdensity in the dark
matter.  According to this model the observed voids today are
primordial negative perturbations that grew gravitationally and
reached shell crossing today.  Shell crossing happens when the radius
of the perturbation has grown by a factor of 1.7, corresponding to a
density contrast of -0.8.  However, at this stage the perturbation is
highly non linear (the corresponding linear amplitude would have been
2.7). Such a large amplitude requires too much power on the scale of
voids and it is inconsistent with the number density of clusters and
super-clusters on slightly lower scales.

In this paper we suggest a new approach that may be used as a
formalism for analyzing voids. We present a simple intuitive model
which describes the formation of voids as due to gravitational growth
and biasing. We relax the assumption that light traces matter on these
large scales and claim that the observed underdensity in the galaxy
density is a product of two factors. The first arises from a simple
gravitational expansion of the negative density perturbation.  The
second factor arises due to biasing: galaxies are less likely to form
in an underdense region. We consider spherical underdensities. This
is, quite generally, a good approximation as negative density
perturbations become more and more spherical as they evolve
(\cite{Icke}, \cite{shu}).  To estimate the biasing factor we use a simple peak
biasing formalism which was developed by \cite{blum} for the
calculation of biasing in clusters.

Our model explains why voids appear in a relatively small range of
sizes and in particular why there is an upper limit to the sizes of
the voids. We use it to calculate the expected sizes and volume
filling factor of voids in different cosmological models and we
compare our results with current observations. The comparison is made
to a simple interpretation of the data - voids occupy 50\% of the
volume, their radii are in the range of 13-30$h^{-1}$ Mpc and the
typical underdensity in the galaxy distribution is taken to be -0.8.
This is a simplified picture and should be modified in the future when
data from new surveys is available and when we have a more refined
model.  Finally we use our model to calculate the expected dark matter
underdensity within the voids.

We find that cosmological models which agree with other constraints
on the power spectrum can in general produce the observed voids even
if not as many as observed. In particular we find, in-spite of the
crudeness of the model and the uncertainties in present day data, that
flat $\Lambda$CDM with a current density parameter
$0.25<\Omega_0<0.35$ is the most preferable model, in agreement with
other observations.

The paper is organized as follows. In section (2) we present the
details of the model and a general calculation of the underdensity of
galaxies inside voids. In section (3) we  calculate the relative volume
of the universe in the form of voids in universes characterized by
different cosmological parameters and CDM power spectra.  We discuss the
implication of our results in section (4).

\section{The Model}

We begin by calculating the dynamics of a negative density
perturbation in a general cosmology. Our goal is to calculate $\eta$,
the ratio of the comoving size of the perturbation to its initial
comoving size, in terms of $\delta_L$, the linear amplitude of the
perturbation. The factor $\eta^{-3}-1$ is the underdensity due to the
gravitational growth. As it is expressed in terms of $\delta_L$ it can
be calculated directly from the linear power spectrum once the
relevant scale is chosen. Then we turn to calculate the underdensity
of galaxies in a larger scale negative density perturbation.

\subsection{Gravitational Growth of Voids}

During the linear phase perturbations grow in amplitude but not in
comoving size. As the perturbations become nonlinear their comoving
radius begins to grow. To find $\eta$ we solve the differential
equation that governs the evolution of a spherical shell surrounding a
negative density region. At some initial time, $t_i$ (at a redshift
$z_i$), the shell is expanding at the same rate as the background
(that is we have an initial density perturbation). The initial small
(negative) density contrast is $\delta_i$ and the initial radius of
the shell is $R_i$.  The background evolution is characterized by the
present values of the Hubble constant $H_0$, the density parameter
$\Omega_0$, and the cosmological constant $\Lambda$. For convenience
we define: $\lambda_0=\Lambda /3H_0^2$.  As long as there is no shell
crossing the mass inside the shell remains constant and energy
conservation yields a differential equation for $R$, the shell's
radius \citep{lahav91}:
\begin{eqnarray}
{\dot{R}}^2= \label{diffr} & & H_0^2[-\Omega_0(1+z_i)^3R_i^2\delta_i
-(\Omega_0+\lambda_0-1)(1+z_i)^2 R_i^2 + \\ & &\nonumber
\Omega_0(1+z_i)^3R_i^3(1+\delta_i)/R+\lambda_0 R^2] .
\end{eqnarray}
We combine this equation with the equation for the background's
redshift:
\begin{equation}
\frac{1}{1+z} \frac{d(1+z)}{dt} = - H_0 P(z)
\end{equation}
where
\begin{equation}
P(z)=\Omega_0(1+z)^3-(\Omega_0+\lambda_0-1)(1+z)^2+\lambda_0
\label{pol}
\end{equation}
to obtain an equation for $dR/dz$. We solve this equation numerically
and obtain the radius $R$ as a function of the redshift.  Since the
comoving radius increases as $1/(1+z)$ we find that $\eta =
[R(z)(1+z)]/[R_i(1+z_i)]$.

The growing solution of the linear perturbation equation is given by
\citet{heath}:
\begin{equation}
\delta_L^+=CP^{1/2}(z)\int_{z}^{\infty}\frac{(u+1)du}{P(u)^{3/2}}.
\label{deltaL}
\end{equation}
The constant C depends on the initial conditions. Since we have
considered earlier an initial density perturbation, $\delta_i$, the
corresponding initial density contrast of the growing mode is
$3\delta_i/5$.

\begin{figure}
\centering
\noindent
\includegraphics[width=10cm]{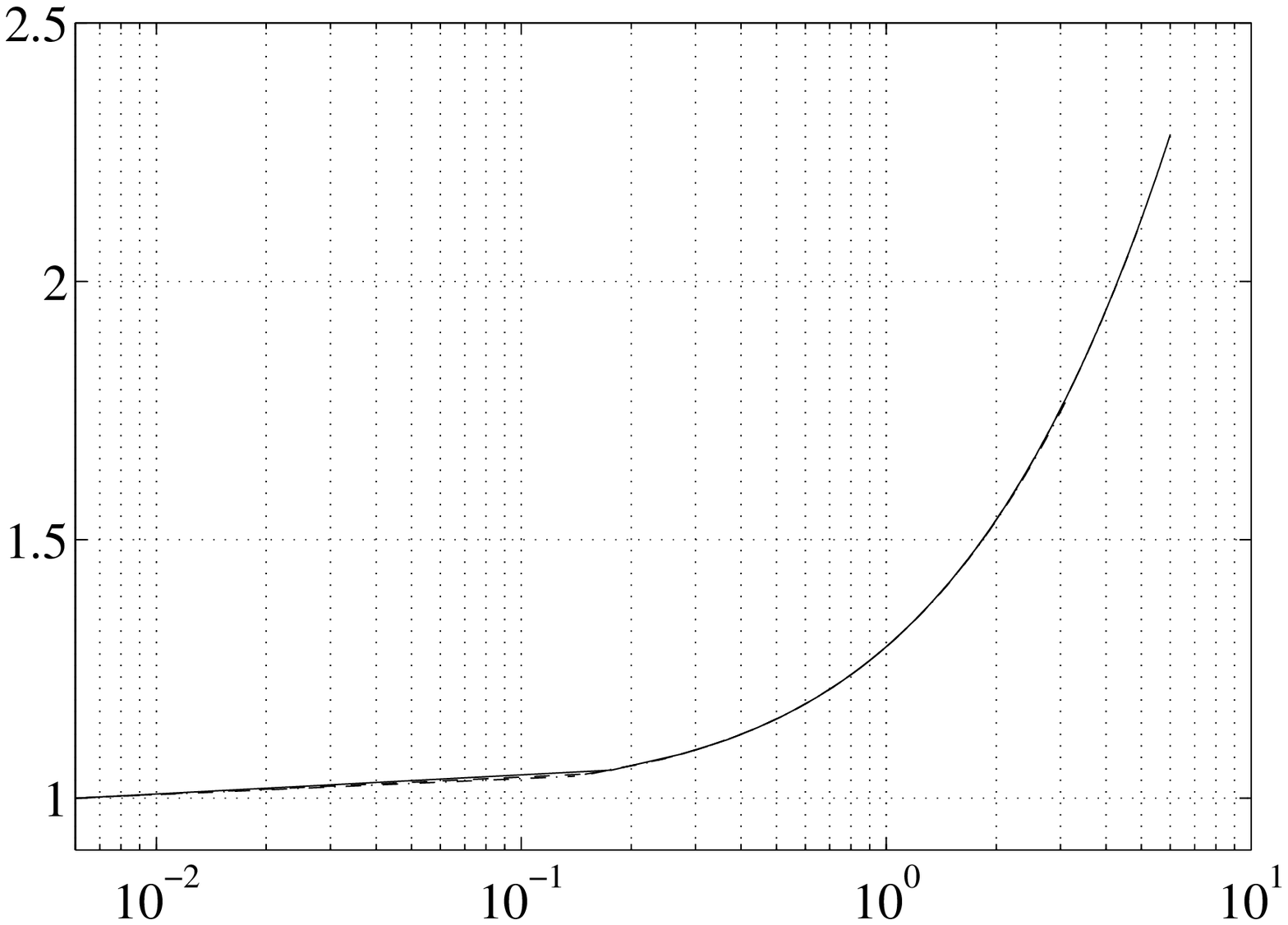}
\caption{\label{growthf}The growth factor, $\eta$, of a spherical
perturbation as a function of the corresponding linear amplitude for
three cases: an $\Omega_0<1$ open universe, $\Omega_0+\lambda_0=1$ and
a flat universe with $\lambda_0=0$.  }
\end{figure}

We can now obtain the growth factor as a function of $\delta_L$ (we
will drop the $^{+}$ from now on). Fig. (\ref{growthf}) depicts $\eta$
as a function of $\delta_L$ for three cases: an Einstein-de Sitter
universe ($\Omega_0=1$, $h=0.5$, $\lambda_0=0$), an open low density
universe without a cosmological constant ($\Omega_0=0.37$,
$\lambda_0=0$) and a flat low density model, with cosmological
constant ($\Omega_0=0.37$, $\lambda_0=0.63$). The function $\eta
(\delta_L)$ is practically independent of the cosmological parameters
and the different curves overlap each other.  For very small
values of $\delta_L$, when the perturbations are still linear, the growth factor
$\eta$ is very close to 1. This is expected since in the linear theory
perturbations grow in amplitude only. $\eta$ increases only as the
perturbation becomes non linear.

\subsection{Biased Galaxy Formation in Voids}

We turn now to the statistical determination of the underdensity of
galaxies within voids. Following \citet{blum} we consider a simple
model in which galaxies form in peaks that exceed a global galaxy
formation threshold. We define the ``efficiency'' of galaxy formation
in some volume $V$, $\epsilon^V$, as the fractional volume of $V$
which is contained in galaxies: \be \epsilon^V=\frac{V_{gal}}{V_{tot}}
\ee

To determine $\epsilon^V$ we use the conditional probability $f(\nu
_{g},\nu _{v})$ of finding a galaxy-size fluctuation with a relative
over-density $\nu_{g}=\frac{\delta _{g}}{\sigma _{g}}$ within a void
size fluctuation with a relative under-density $\nu
_{v}=\frac{\delta_v}{\sigma_{v}}$. Here $\sigma_g$ and $\sigma_v$
are the rms mass fluctuations filtered on galaxy and void scales,
respectively. The scale of a galaxy, $R_g$ is related to its mass
$M_g$ (including the dark matter) through $M_g=(4\pi/3)R_g^3<\rho>$.

\begin{equation}
f(\nu _{g},\nu _{v})=\frac{1}{2\pi \sqrt{1-r^{2}}}\exp
[-\frac{1}{2\sqrt{ 1-r^{2}}}(\nu _{g}^{2}+\nu _{v}^{2}+2r\nu _{g}\nu
_{v})].
\label{biv}
\end{equation}
$r$ is the correlation coefficient between the two scales, given by
$\sigma_{gv}^{2}/\sigma_{g}\sigma _{v}$, where:
\begin{equation}
\sigma _{gv}^{2}=\frac{1}{(2\pi )^{3}}\int d^{3}\mathbf{k}\left|
\delta _{ \mathbf{k}}\right| ^{2}W(kR_{g})W(kR_{v}),
\label{corr}
\end{equation}
and $W(kR)$ is a window function. We choose a top-hat window
function:
\begin{equation}
W(kR)=\frac{3(\sin{kR}-kR\cos{kR})}{(kR)^3}
\label{win}
\end{equation}

We use a single typical galaxy mass of $M_g=1.2\cdot10^{12}
M_{\odot}\frac{(M/L)}{100M_{\odot}/L_{\odot}}$ which is the median of
the galaxy luminosity function. This is clearly an approximation and
possibly the crudest one we make in this work.  The scale related to
this mass, $R_g$, varies according to the cosmological parameters of
the model.  Assuming that only the peaks that exceed a global
threshold $\nu _{th}$ become luminous galaxies the efficiency of
galaxy formation in voids of a given radius $R$ and with a given
$\nu _{v}$ is:
\begin{equation}
\epsilon^{void} (\nu _{v},R)=\frac{\int_{\nu _{th}}^{\infty }f(\nu
_{g},\nu _{v})d\nu _{g}}{\int_{-\infty }^{\infty }f(\nu _{g},\nu
_{v})d\nu _{g}} \label{reff} =\frac{1}{2}{\rm erfc}\left[ \frac{\nu
_{th}+r\nu _{v}}{\sqrt{2(1-r^{2})}}\right] .  \label{erfc}
\end{equation}
The galaxy formation threshold, $\nu_{th}$, is calculated using the
global efficiency of galaxy formation: \be
\epsilon^{bg}=\epsilon^{void}(R_v=\infty)=\frac{1}{2} {\rm
erfc}\frac{\nu_{th}}{\sqrt{2}} \ee

Empirically, one possible way of determining the fraction of mass
residing in galaxies is to divide the mass-to-light ratio of a typical
galaxy by the mass-to-light ratio of the universe. Following
\citet{bah95} we take the M/L ratio of the universe to be
1350$\Omega_0$h and that of a typical galaxy to be 100h. Now we have
\begin{equation}
\frac{1}{2} {\rm
erfc}\frac{\nu_{th}}{\sqrt{2}}=\frac{(M/L)_{gal}}{1350\Omega_0 h}
\label{staerf}
\end{equation}
Using eq. \ref{staerf} we can determine, for any given $\Omega_0$, the
global galaxy formation threshold.

\subsection{The combined  underdensity}
The current underdensity of galaxies in voids, $\delta_{gal}$, is \be
1+\delta_{gal}=\frac{\rho_{gal}^{void}}{\rho_{gal}^{bg}}=
\frac{\epsilon^{void}}{\epsilon^{bg}\eta^3} \ee where
$\rho_{gal}^{void}$ is the density of galaxies in the void and
$\rho_{gal}^{bg}$ is their average density in the background
universe. The second equality holds once the growth factor of the void
is taken into account and all the galaxies are taken to be of the same
typical scale ($\approx 1 h^{-1}$Mpc).

\section{The Void Content of the Universe in Different CDM Models}

Given the model described above we shall now calculate the
expected sizes and the volume filling factors of voids in different
cosmological models. We also calculate the dark matter underdensity in
voids in these models. Our aim is to find the
dependency of the filling factor on cosmological parameters. A second goal is to predict the dark matter underdensity in voids and through this to learn of the biasing between dark and luminous matter on these large scales.

We consider first the SCDM model ($\Omega_0=1$, $h_0=0.5$,
$\Lambda=0$, $n=1$, $\Omega_b=0.0125h^{-2}$). It is already
established that this is not a valid model of the universe; it
does not agree simultaneously with COBE and with cluster abundance
data. However because of the simplicity of the SCDM model we use it as a tool to demonstrate how the
void content of the universe changes with the normalization of the
power spectrum. We use the transfer function calculated by
\cite{Bardeen} as the shape of the dark matter  power spectrum. For
the normalization we consider two possibilities: COBE normalization
as calculated by \cite{bunn}  ($\sigma_8=1.27$) and cluster abundance
normalization as given by \cite{pen} ($\sigma_8=0.53$).

\begin{figure}
\centering
\noindent
\includegraphics[width=9cm]{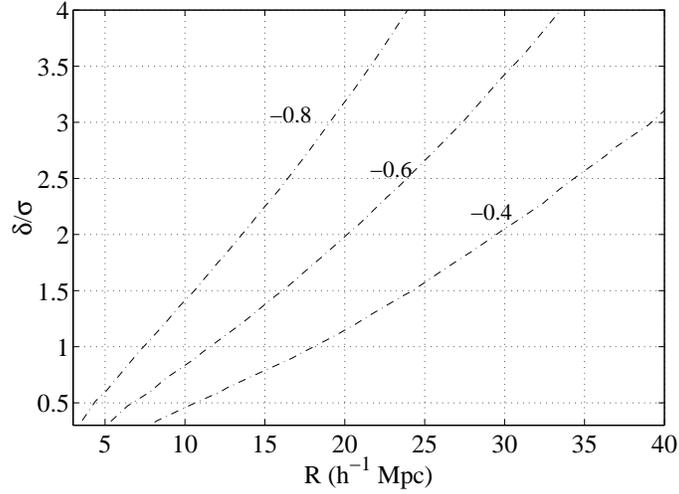}
\caption{\label{contours_scdm2}Contours of constant $\delta_{gal}$ are
displayed as a function of the radius of the voids today, R, and the relative underdensity of the dark matter, $\nu_v$,
in a standard CDM model which is cluster-normalised. We can see that the typical voids ($R=20
h^{-1}$Mpc $\delta_{gal}=-0.8$) are 3$\sigma$ objects i.e they are
produced very rarely.  }
\end{figure}

We present contour lines of constant galaxy underdensity as a function
of the radius of the voids today $R$ and the relative underdensity in
the dark matter $\nu$. Figure [\ref{contours_scdm2}] depicts several
contour lines for the cluster normalized SCDM model. If we look at
constant $\nu$ we find that there are relatively more galaxies in
larger voids. This is due in part to the statistical properties of the
fluctuations and in part to the gravitational expansion of the
underdensities. At larger scales the amplitude of the perturbations is
smaller, so to form a galaxy in a larger underdensity we need galaxy
size perturbations of smaller amplitude. These will be more abundant
because the distribution function of the fluctuations is a Gaussian.
Thus there will be more galaxies in larger voids and the relative
underdensity of the galaxies will decrease. The gravitational
expansion factor does not compensate for this - in fact, it becomes
less important because  $\delta_L$ of the underdensities
decreases as $R$grows (see Fig[\ref{growthf}]).  If we look at voids of constant
radius we see that there are relatively less galaxies at larger $\nu$.
The contribution to this behavior is also two-fold. Negative
perturbations of higher $\nu$ correspond to deeper voids; In such
voids we need galaxy size perturbations of larger amplitude to form
galaxies. These are less abundant and therefore the relative
underdensity of the galaxies is larger in deeper voids. To this we add the fact that
negative perturbations of higher $\nu$ are of higher $\delta_L$ and
for these the gravitational factor is bigger. Thus the volume of the
void will grow and the relative underdensity in the galaxy
distribution will be even greater.

An important feature to notice in this figure is that for a given
underdensity of the galaxy distribution inside a void, larger voids
are produced exponentially more rarely as they require large and
hence extremely rare initial perturbations.  Thus there is a sharp
upper limit to the sizes of voids.

To compare with observations we calculate the filling factor of the
voids. We calculate the fraction of the universe which is composed of
spherical and isolated negative density perturbations that are large
enough and deep enough to produce voids of radii 13-30 $h^{-1}$Mpc.
The number of spherical inhomogeneities of a radius R and an amplitude in
the range [$\delta,\delta+d\delta$] inside the horizon is:
\begin{equation}
N(R,\delta)d\delta=\frac{c^3H_0^{-3}}{R^3}
\cdot\frac{1}{(2\pi\sigma_R^2)^{1/2}}e^{-\delta^2/2\sigma_R^2}d\delta.
\end{equation}
Clearly the isolated spherical approximation would break down at low
$\nu$ values and it might be violated around the lower limits ($\nu
\sim 1.5$) of our integration. We expect it to hold at higher values.
The total volume of the corresponding voids is $(\eta R)^3\cdot
N(R,\delta)$ and the relative volume is
\begin{equation}
f(R,\delta)d\delta=\frac{\eta^3}{(2\pi\sigma_R^2)^{1/2}} \cdot
e^{-\delta^2/2\sigma_R^2}d\delta
\label{fill}
\end{equation}

The voids with radii 13-30 $h^{-1}$Mpc correspond to initial
fluctuations of sizes of about 10-25$h^{-1}$Mpc, depending on the
model.  Thus to obtain the overall filling factor, denoted by $f$, we
integrate equation (\ref{fill}) along the contour of
$\delta_{gal}=-0.8$ in the appropriate range. This method of counting
might be complicated by the possibility of over-counting: a void of certain radius and amplitude might be counted again as a void of larger radius and
smaller amplitude. This cannot happen if the underdensity $\delta$
increases with R, as the larger void would be deeper. We therefore checked the behavior of  $\delta$, and found that it increases monotonically with R.

\begin{figure}
\centering
\noindent
\includegraphics[width=9cm]{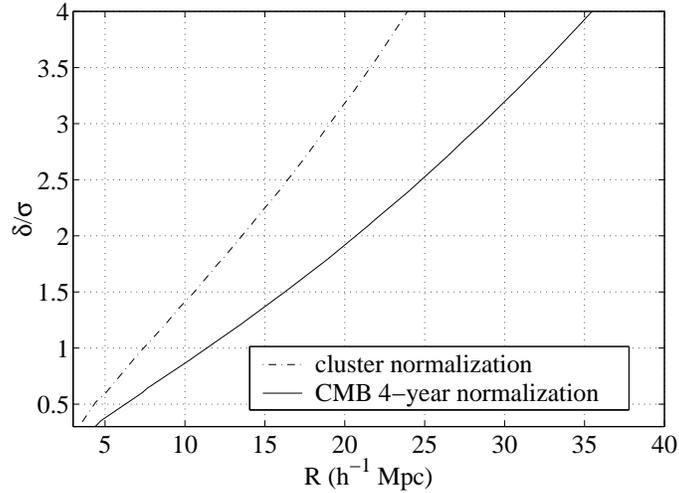}
\caption{\label{2scdm}Two contours of $\delta_{gal}=-0.8$ are
displayed. One is for cluster abundance normalized SCDM
(dashed). The other is for COBE-normalized SCDM. The effect of the different
normalizations is very obvious: When the power on the scale of voids is
higher (COBE) the observed voids are much more frequent.  }
\end{figure}

We have carried out this calculation for two CDM models with different normalizations. The contour lines
of $\delta_{gal}=-0.8$ of the two models are presented in Fig. [\ref{2scdm}].
The difference between the two models is very pronounced: the model with
more power on the scale of the voids (COBE normalized) yields more voids.  This can be explained as follows: when
the power on the scale of the voids is larger, the amplitude
needed to produce the voids that we see today is reached by fluctuations
with lower $\nu$ which are therefore more frequent. In models
with less power on void scales, the same amplitude of underdensities
requires higher $\nu$ values and are therefore less frequent.
This is reflected in the calculated values of the filling factors
$f=33\%$ for the COBE  normalized model, and only 11\% for the cluster
normalized model.

However, neither of these CDM models is an acceptable model of the
universe.  We estimate now the void content of the universe in the
context of power spectra which are compatible with observations. We
consider Open CDM and flat $\Lambda$CDM models.  The transfer function
used is, as above, \citep{Bardeen}. The normalization will be
according to the 4-year COBE DMR experiment, as calculated by
\cite{bunn}.  To determine the models' parameters we first set
$\Omega_0$, then we choose a tilt such that the model is also cluster
normalized. This is done by calculating $\sigma_8$ and finding a tilt
such that the normalization condition given by \cite{pen}:

\begin{equation}
\sigma_8=(0.53\pm0.05)\Omega_0^{-0.45}
\end{equation}
for open models and
\begin{equation}
\sigma_8=(0.53\pm0.05)\Omega_0^{-0.53}
\end{equation}
for flat models is satisfied. We take Hubble's constant to be $H_0=65$
 $Km sec^{-1}Mpc^{-1}$ in agreement with recent results from HST Key
project ($H_0=0.71\pm0.06$, see \cite{mould}) and measurements of time
delay between multiple images of gravitational lens systems
($H_0=0.69^{+13}_{-19}$, see \cite{biggs}). $\Omega_bh^{-2}=0.015$ in
all models. The models are described in Table 1.  The first two
columns of the table give $\Omega_0$ and $n$. In the third column we
list $\sigma_8$, the amplitude of mass fluctuations in spheres of
radius 8$h^{-1}$Mpc. All $\sigma_8$ are within the ranges allowed by
the cluster normalisation. As another check for the validity of our
models, we show that the shape parameter $\Gamma$ of each of the
models is within limits ($0.15<\Gamma<0.3$) allowed by measurments of
the angular correlation function from the APM galaxy survey
\citep{ebw}. The values of $\Gamma$ are listed in the fourth
column. Finally the fifth column gives the calculated filling factor and
in the sixth column the calculated dark matter underdensity is listed.

\begin{figure}

\centering
\noindent
\includegraphics[width=9cm]{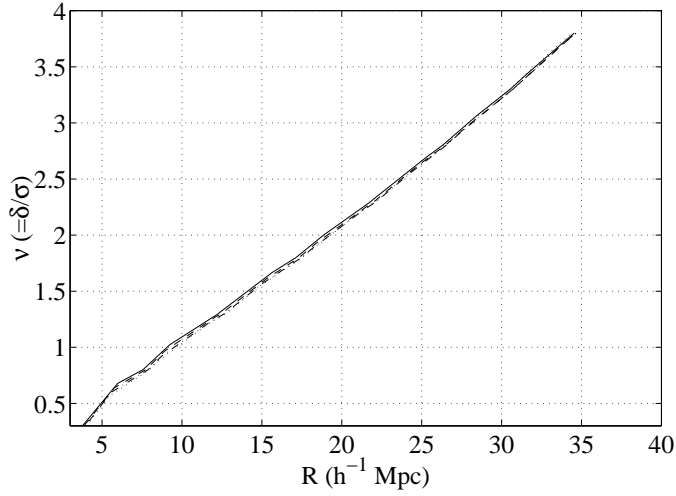}
\caption{\label{con_in_open_cdm}Contours of $\delta_{gal}=-0.8$ in the
four Open CDM models we checked are displayed as a function of the
radius of the voids today, R, and relative underdensity, $\nu$. It is
clear that in open models the distribution of the voids is not a
function of $\Omega_0$. This can be  explained by the fact that the
power spectra for these models are almost the same on all scales.  }
\end{figure}

\begin{figure}
\centering
\noindent
\includegraphics[width=10cm]{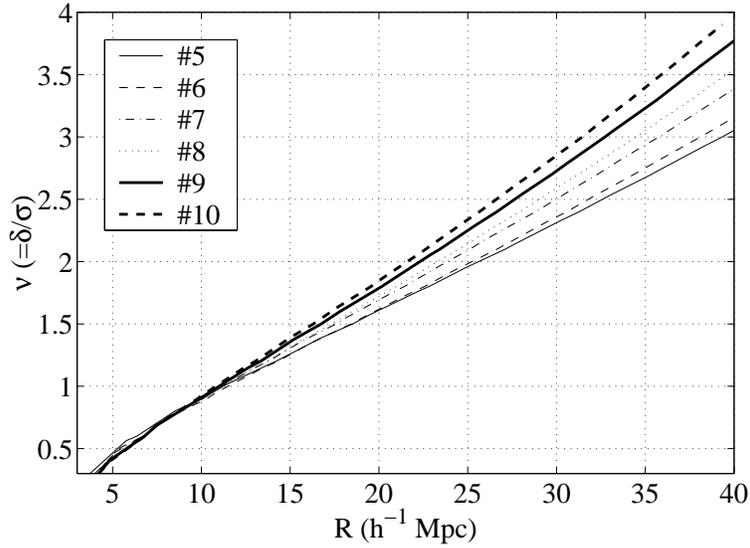}
\caption{\label{con_in_flat_cdm}The contours of $\delta_{gal}(\nu,
  R)=-0.8$ in the six $\Lambda$CDM models as a function of the
  radius of the voids today, $R$, and of the relative underdensity of
  the void, $\nu$. Here we see a stronger dependance on $\Omega_0$, and
  also we see that voids are more frequent in these models than in the
  open models.  }
\end{figure}

As before, we present the results as contour lines of constant
$\delta_{gal}$ as
a function of the radius of the voids today, $R$, and as a function
of their relative underdensity in the dark matter, $\nu$. The contour lines
for the Open and $\Lambda$ CDM models are presented in figures
[\ref{con_in_open_cdm},\ref{con_in_flat_cdm}] respectively. We notice,
first, that all the models show a common behaviour which
was manifested also in the SCDM models: Larger voids of
$\delta_{gal}=-0.8$ are produced exponentially more rarely. The sharp
upper limit to the sizes of voids exists in all CDM models.

\begin{table}
\begin{tabular}{|c|l|c|c|c|c|c|l|} \hline
& &{\boldmath $\Omega_0$}&{\boldmath n}&{\boldmath
$\sigma_8$}&{\boldmath $\Gamma$}&{\boldmath $f(\%)$}&{\boldmath
$\delta_{DM}$ } \\ \hline
1& {\bf Open-} &0.3&1.3&0.92&0.18&19&-0.56\\ \cline{3-8}
2&{\bf CDM}&0.35&1.17&0.85&0.21& 18&-0.53\\ \cline{3-8} 3&
&0.4 &1.07 &0.81&0.24& 18 &-0.52\\ \cline{3-8}
4& &0.45&0.98&0.76&0.27&18&-0.49\\ \hline
5&{\bf $\Lambda$CDM}&0.2&1.2&1.2&0.11&31&-0.60 \\ \cline{3-8}
6&{\bf(flat)}&0.25&1.1&1.11&0.15&29&-0.56 \\ \cline{3-8}
7&&0.3&1&0.95&0.18&25&-0.53\\ \cline{3-8}
8& &0.35&0.96 &0.93&0.21 &25&-0.52\\\cline{3-8}
9& &0.4 &0.91 &0.86&0.24 &24&-0.50\\ \cline{3-8}
10& &0.45&0.88&0.83&0.27 &22 &-0.49\\ \hline

\end{tabular}
\caption{\label{table}A list of the models considered.}
\end{table}

Figure [\ref{con_in_open_cdm}] describes voids of $\delta_{gal}=-0.8$
in Open CDM models. It is clear that the void distribution does not
depend strongly on $\Omega_0$. The filling factor is almost constant, having values 18-19\%.  Figure
[\ref{con_in_flat_cdm}] describes the same voids in flat $\Lambda$CDM
models. Here there is a stronger dependence of the void distribution
function on $\Omega_0$, and the filling factor is larger than in the
open models. It is in the range 22-31\%, decreasing with
$\Omega_0$.

We have also calculated the expected underdensity of the dark matter.
This underdensity is given simply by $\delta_{DM}=\eta^{-3}-1$. It is
listed, for 20 $h^{-1}$Mpc voids in the different models, on sixth
coloumn of table [1]. The underdensities are in the range [-0.5,-0.6]
for all the models. These typical values are a factor of 1.3-1.6
smaller than the galaxy underdesity, indicating this factor as the
biasing between galaxies and dark matter perturbations on the $20
h^{-1}$Mpc scale within the voids.

\section{Discussion}

We have presented here a model for the formation of voids.  In this model 
voids
arise from initial negative density perturbations. Such underdensities
grow in comoving volume and this growth increases the underdensities
of both the galaxies and the dark matter within the voids. The galaxy
underdensity is enhanced further since positive galaxy size
perturbations are less frequent within negative void size
perturbations.  This mechanism inhibits the formation of
galaxies within the voids.  In our model both mechanisms contribute
comparable factors to the overall galaxy underdensity.

We use the model to investigate the void content of the universe
for different power spectra which are in agreement with COBE and
cluster abundance data. Qualitatively we found,
in all the cosmological models we tested, that the probability of finding voids of a
certain $\delta_{gal}$ falls exponentially with the radius. This
behavior may explain the observed upper limit of the radii of voids.

In order to quantitatively test our model, we have calculated the
filling factor of the observed voids ($R_v\in[13-30]h^{-1}$Mpc,
$\delta_{gal}=-0.8$). We find that in all the models that we considered the
observed voids fill only half of the expected volume. However,
there is a clear trend toward higher filling factors in $\Lambda$CDM
models where the relevant voids appear more frequently and fill a
larger fraction of the universe. We also found that in the open models
that we have tried, since the power spectra were very similar, the
distribution of the void sizes and the filling factor did not change
with $\Omega_0$. However, in $\Lambda$CDM models, as $\Omega_0$ grows the
relevant voids become less frequent and the void content of the
universe decreases. It can be explained by the
fact that as we increase $\Omega_0$, the amplitude of fluctuations on
the scale of voids is decreased.  The most preferable models are the
$\Lambda$CDM models with $0.25<\Omega_0<0.35$: these comply with all the constraints and have the highest void filling factors.   Still even these values fall short of the observations by a factor of $\approx1.4$.

We suspect that the small
filling factor is due in part to the oversimplified model of galaxy
formation that we have used. A more realistic model should allow for a range
of galaxy masses and a more elaborate biasing mechanism between the dark
matter and galaxies. This will be the next step towards a more reliable model.
Also note that, as already mentioned, another important
assumption of our model is that of  spherically symmetric isolated evolution.
We have assumed
that the underdensities are spherical and isolated when calculating
the gravitational growth and the filling factor, ignoring
possible mergers between neighboring voids and the influence of positive
overdensities on nearby underdensities. Void
mergers might lead  to the disappearance of smaller voids
with deeper underdensities alongside with the appearance of
larger asymmetric voids.
Positive nearby overdensities could exert forces on matter inside
underdensities and increase their growth rate.
Both effects could increase the filling factor of voids.

Finally we have computed the underdensity of dark matter in typical
voids of radius 20$h^{-1}$Mpc. While the dark matter is influenced
only by the gravitational expansion of the negative density
perturbations the number of galaxies is also influenced by the biasing
factor. For this reason we have $\mid
\delta_{DM}\mid<\mid\delta_{gal}\mid$.  The expected
dark matter underdensities that we find are about a factor of
1.3-1.6 lower than the underdensities of the galaxy density. These values
should be regarded only as an upper limit to the
real underdensity expected in nature. Since real voids are more
frequent, they must correspond to lower $\nu$ values and their gravitational
growth factor would be smaller.  This will result in a less negative
dark matter density contrast. This prediction should be compared
with estimates of the dark matter density in voids from N-body
simulations and with future measurements of the dark matter
underdensity within the voids.

It will be interesting to apply our model to account for evolution of
void sizes and abundances as a function of redshift. We suspect that
in critical density universes the evolution of voids will be stronger
than in low density universes: We have shown in section 2 that the growth
of the radius of the void depends only on the linear amplitude of the
perturbation and not on cosmological parameters. Thus in a universe
with $\Omega_0=1$ where the linear amplitude grows like the scale factor,
the radius of the void will grow constantly. However, in models  where
matter ceases to dominate, such as open models which become curvature
dominated at small z or flat models with a cosmological constant which
begins to dominate at late times the linear amplitude reaches a constant
value and stops growing. In such cases, the comoving radius of the voids
will also stop growing at late times. Thus we could use the model to predict
the change in comoving radius of voids as function of z in different
cosmological models and by comparing to the next generation of deep
sky surveys discriminate between low and critical density models
(for example, in critical density universes older voids will be
smaller in radius and the galaxy underdensity in them will also be smaller).

Upcoming sky surveys, such as the Sloan Digital Sky Survey, will
increase the available galaxy distribution data by several orders of magnitude. In particular such surveys will include more voids and
hopefully enough voids to obtain the distribution and
evolution of the void sizes.  That would allow us to
compare our model with observations in a more accurate way and
constrain $\Omega_0$ and other cosmological parameters  using the void
distribution.

\acknowledgements We thank G. R Blumenthal for helpful discussions at
the beginning of our work. We also thank H. El-Ad, A. Dekel, T. Kolatt and V. De Lapparent for their usefull comments. Y. F. wishes to thank IAP and D.A.R.C 
for the warm hospitality during the last stages of this work.



\end{document}